\documentclass[10pt,a4paper]{iopart}
\usepackage{iopams}
\usepackage{amsopn}
\usepackage{mathabx}
\usepackage{bm}
\usepackage{mathptmx}           
\usepackage{graphicx}
\usepackage{comment}
\usepackage[dvipsnames]{xcolor}
\usepackage{cite}
\usepackage[colorlinks=true,allcolors=blue,bookmarks=true,bookmarksnumbered=true]{hyperref}

\DeclareMathOperator{\Ci}{Ci}

\newcommand{\eqref}[1]{(\ref{#1})}
\newcommand*{\tran}{^{\mkern-1.5mu\mathsf{T}}}

\begin{document}

\title[]{Noise-induced servo errors in optical clocks utilizing Rabi interrogation}

\author{T Lindvall, A E Wallin, K J Hanhij\"arvi and T Fordell}

\address{VTT Technical Research Centre of Finland Ltd, National Metrology Institute VTT MIKES, P.O.\ Box 1000, FI-02044 VTT, Finland}
\ead{thomas.lindvall@vtt.fi}

\begin{abstract}
We show that in optical clocks based on Rabi interrogation, both laser-frequency and magnetic-field flicker ($1/f$) noise with zero mean can lead to servo errors at the $10^{-18}$ level if the negative-detuning (red) and positive-detuning (blue) sides of the transition are always probed in the same order. This is due to the strong correlations of flicker noise in combination with an imbalance in the response of the servo discriminator to positive and negative differential frequency noise between the red- and blue-side probing. This imbalance is particularly large for a normalized discriminator.
We derive an analytical expression for the servo error based on the correlation function of the laser-frequency or magnetic-field noise and compare it to numerical servo simulations to demonstrate how the error depends on the noise level, servo parameters, and probing sequence. We also show that the servo error can be avoided by normalizing the discriminator with a moving mean or by reversing the red/blue probing order for every second servo cycle.
\end{abstract}

%
\vspace{2pc}

\vspace{2pc}
\noindent Accepted to Metrologia. Published version will be available Open Access.\\
\noindent \href{https://doi.org/10.1088/1681-7575/acdfd4}{DOI 10.1088/1681-7575/acdfd4}.
%
%
%
\ioptwocol

\section{Introduction \label{sec:intro}}

Today, optical atomic clocks based on both single ions in radiofrequency traps \cite{Brewer2019a,Sanner2019a} and neutral atoms in optical lattices \cite{McGrew2018a,Bothwell2019a} report estimated fractional systematic uncertainties of $1\times 10^{-18}$ to $3\times 10^{-18}$.
To reach this level of uncertainty, one must consider all possible frequency shift mechanisms,
also those of technical nature, such as servo errors in the frequency lock to the atoms or ion.

Servo errors caused by laser-frequency drift \cite{Peik2006a,Falke2011a,Dube2018a,Yuan2021a} and magnetic-field drift \cite{Barwood2015a,King2022a,Steinel2022a} have been studied. It is also well known that laser \cite{Leroux2017a} and magnetic-field~\cite{Barwood2015a} noise can degrade the stability of an optical clock. Laser noise can also do this through the Dick effect, i.e., periodical sampling of the noise due to dead time in the clock cycle~\cite{Dick1987a,Santarelli1998a}.
The frequency bias caused by noise has been considered in the context of Cs beam standards using Ramsey interrogation \cite{Shirley2001a}, but the error signal was linearized, which leads to the conclusion that noise with a zero mean causes no bias.

Cavity-stabilized clock lasers are fundamentally limited by the thermal flicker ($1/f$) noise floor, and flicker noise is often found also in the magnetic field. Here, we show that this noise can lead to servo errors at the $10^{-18}$ level in clocks based on Rabi interrogation if the negative-detuning (red) and positive-detuning (blue) sides of the transition are always interrogated in the same order.
In the next section, the model for the servo discriminator is introduced.
In Section~\ref{sec:se}, analytical expressions for the noise-induced servo error are derived. Section~\ref{sec:noise} discusses laser and magnetic-field noise, its correlation function and the effect of dead time. Numerical servo simulations are described in Section~\ref{sec:sim} and Section~\ref{sec:results} then presents numerical results comparing the analytical and numerical approaches. Finally, Section~\ref{sec:summary} discusses how to avoid the noise-induced errors and summarizes the results.
The results are given for parameters typical for single-ion clocks, but can be adapted to other types of clocks by updating the numerical values.
Although the paper concentrates on Rabi interrogation, a brief treatment of Ramsey interrogation, for which the noise-induced error is found to be negligible, is presented in \ref{app:Ramsey}.

\section{Model \label{sec:model}}

We consider an optical atomic clock where the atoms are interrogated with Rabi spectroscopy, i.e., using single pulses of laser radiation incident on the atoms, typically at a constant power, for a duration $\tau$. To obtain an error signal, the transition is probed with equal positive and negative detunings from the estimated line center, typically close to the expected half-width at half-maximum (HWHM) points. The clock may operate on a single magnetically insensitive transition or multiple transitions may be probed to cancel the linear Zeeman shift and possibly also tensor shifts.

\begin{figure}[tb]
\centering
\includegraphics[width=0.85\columnwidth]{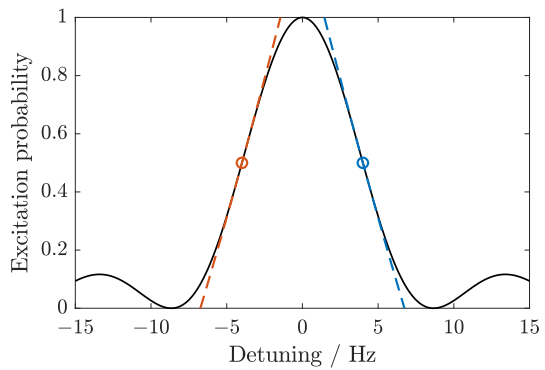}
\caption{Rabi lineshape for 100-ms $\pi$-pulses (solid black) and second-order series expansions (dashed) around the red and blue half-maximum points (circles).
\label{fig:lineshape}}
\end{figure}

When clock servos are modelled, the atomic lineshape is often linearized around the nominal probing points. To analyze noise-induced servo errors, at least a second-order expansion is required.
In this case, the mean excitation probabilities for the blue (B) and red (R) sides of the lineshape can be written as
\begin{equation} \label{eq:pBR}
p_{\mathrm{B,R}} = p_0 ( 1 \mp a\, \delta'_{\mathrm{B,R}} + b\, \delta^{\prime 2}_{\mathrm{B,R}}).
\end{equation}
Here $p_0$ is the mean excitation probability at the nominal probing points, which affects the quantum projection noise (QPN) but not the servo errors considered here, and $\delta'_{\mathrm{B,R}} = \delta + \delta_{\mathrm{B,R}}$, where $\delta$ is the systematic laser detuning error from the resonance and $\delta_{\mathrm{B,R}}$ is the laser noise (angular frequency) during the interrogation of the blue or red side.\footnote{Formally, the expansion is done in the dimensionless detuning $\delta\tau$, but $\tau$ and $\tau^2$ are absorbed into the coefficients $a$ and $b$ to simplify the notation.}
In general, the lineshape and thus the parameters $a$ and $b$ will depend on the atom/ion temperature, the lifetime of the clock state, the pulse area (duration and amplitude), and on possible broadening by laser and magnetic-field noise.
For simplicity, the numerical results will be given for an ideal Rabi lineshape with the excitation probability
\begin{equation} \label{eq:rho_ee0}
p_\mathrm{e}(\delta_0) = \frac{\Omega_0^2}{2(\Omega_0^2 + \delta_0^2)} \left[1-\cos{\left(\sqrt{\Omega_0^2 + \delta_0^2} \tau\right)}\right],
\end{equation}
where $\Omega_0$ is the Rabi frequency and $\delta_0$ the laser detuning from the resonance,
and perfect $\pi$-pulses (pulse area $\Omega_0 \tau = \pi$).
This gives $a=0.604 \tau$ and $b = 0.0141 \tau^2$ for expansion around the HWHM points of $\pm 2.51/\tau$ ($\pm0.399\;\mathrm{Hz}/(\tau/\mathrm{s})$ in frequency units). Figure~\ref{fig:lineshape} illustrates this for a pulse length of 100\;ms. The analytical results can, however, be applied to any lineshape that can be numerically expanded as \eqref{eq:pBR}.

In the simplest case, the red and blue sides of a transition are probed only once per servo cycle (but usually with multiple pulses in single-ion clocks) and in a fixed order, which we refer to as RB or BR probing depending on the order. In general, each side can be probed multiple times per cycle, e.g., in the order RBBR or using higher-order Thue--Morse sequences \cite{Schat2007a} in order to cancel the error caused by laser amplitude drift \cite{Itano2007a}.
The number of successful excitations are then summed, corresponding to averaging the transition probabilities \eqref{eq:pBR} over all B (R) interrogations during a cycle.

Two different servo discriminators are commonly used in clocks using Rabi interrogation. The normalized discriminator, see, e.g., \cite{Dube2015a, Barwood2015a,King2012a,Yuan2021a}, can be written as
\begin{equation} \label{eq:Dn}
D_{\mathrm{norm},n} = \frac{p_{\mathrm{B},n}-p_{\mathrm{R},n}}
{p_{\mathrm{B},n}+p_{\mathrm{R},n}},
\end{equation}
while the unnormalized one, see, e.g., \cite{Peik2006a,Holliman2022a,Steinel2022a}, can be written as
\begin{equation} \label{eq:Du}
D_{\mathrm{un},n} = \frac{p_{\mathrm{B},n}-p_{\mathrm{R},n}}{2 p_0}.
\end{equation}
Here $n$ is the servo-cycle number. The normalized discriminator \eqref{eq:Dn} makes the applied frequency correction immune to slow changes in the excitation probability, but, as will be seen later, results in significantly larger noise-induced servo errors.
In single-ion clocks, the excitation probabilities $p_{\mathrm{B,R},n}$ can be very noisy even in the absence of laser and magnetic-field noise due to QPN. However, the QPN averages down statistically and will not be included in the analytical model.

To study noise-induced servo errors, it is sufficient to consider a simple integrating servo for locking to an atomic transition; the effect of laser-drift compensation will be considered separately in Section~\ref{sec:noise}.
During the servo cycle $n$, the atomic transition is probed with a frequency $\omega_n(t) = \omega_\mathrm{las}(t) + \omega_{\mathrm{s},n-1}$ (the stepping to the R and B sides is implicitly assumed). Here, $\omega_\mathrm{las}(t)$ is the `free-running' (i.e., not stabilized to the transition) laser frequency, which in general includes both drift and noise, and $\omega_{\mathrm{s},n-1}$ is the servo frequency determined at the end of the previous cycle, which is updated using the recursion formula $\omega_{\mathrm{s},n} = \omega_{\mathrm{s},n-1} + GD_{i,n}$, where $G$ is the gain and $i= \{\mathrm{norm}, \mathrm{un}\}$.

To study noise-induced effects, the detuning from the transition frequency $\omega_0$ during cycle $n$ is evaluated separately at the time $t_\mathrm{B,R}$ of the B and R interrogations, as in \eqref{eq:pBR}, and is separated into a systematic detuning error and laser noise, $\delta'_{\mathrm{B,R},n} = \omega_n(t_\mathrm{B,R}) -\omega_0 = \delta_{n-1} + \delta_{\mathrm{B,R},n}$. If all time dependence of the laser frequency $\omega_\mathrm{las}(t)$ is contained in the noise terms $\delta_{\mathrm{B,R},n}$, the systematic detuning error will obey the same recursion formula as the servo frequency, $\delta_{n} = \delta_{n-1} +  GD_{i,n}$.
Taking into account that $p_{\mathrm{B,R},n}$ can be the sum of multiple expressions like \eqref{eq:pBR}, the second-order expansion of the normalized discriminator then becomes
\begin{eqnarray}
D_{\mathrm{norm},n} = & -a \left[ \delta_{n-1} + \frac{\bar{\delta}_{\mathrm{B},n} + \bar{\delta}_{\mathrm{R},n}}{2} \right] \left[ 1 + \frac{a}{2} (\bar{\delta}_{\mathrm{B},n} -\bar{\delta}_{\mathrm{R},n}) \right] \nonumber \\
& +b \left[ \delta_{n-1} (\bar{\delta}_{\mathrm{B},n} -\bar{\delta}_{\mathrm{R},n})
+ \frac{\widebar{\delta_{\mathrm{B},n}^2}}{2} - \frac{\widebar{\delta_{\mathrm{R},n}^2}}{2} \right]. \label{eq:Dns}
\end{eqnarray}
Here, the bars denote the mean over all interrogations on the red or blue side during the servo cycle.
The unnormalized discriminator is otherwise identical, but the $a^2$ term is missing,
\begin{eqnarray}
D_{\mathrm{un},n} = & -a \left[ \delta_{n-1} + \frac{\bar{\delta}_{\mathrm{B},n} + \bar{\delta}_{\mathrm{R},n}}{2} \right]  \nonumber \\
& +b \left[ \delta_{n-1} (\bar{\delta}_{\mathrm{B},n} -\bar{\delta}_{\mathrm{R},n})
+ \frac{\widebar{\delta_{\mathrm{B},n}^2}}{2} - \frac{\widebar{\delta_{\mathrm{R},n}^2}}{2} \right]. \label{eq:Dus}
\end{eqnarray}
In (\ref{eq:Dns}--\ref{eq:Dus}), the first term is the linear discriminator term desired for tracking the transition frequency. The dependence on the differential noise, $\bar{\delta}_{\mathrm{B},n} -\bar{\delta}_{\mathrm{R},n}$, is in the unnormalized discriminator \eqref{eq:Dus} due only to the curvature $b$ of the lineshape, while it is dominated by the significantly larger $a^2$ cross term between the numerator and denominator in the normalized discriminator \eqref{eq:Dns}.

In the special case of RB or BR probing, each side is probed only once per cycle and thus $\widebar{\delta_{\mathrm{R,B},n}^2} = \bar{\delta}_{\mathrm{R,B},n}^2 = \delta_{\mathrm{R,B},n}^2$, so that  (\ref{eq:Dns}--\ref{eq:Dus}) simplify to
\begin{equation} \label{eq:Dis}
D_{i,n} = -a \left[ \delta_{n-1} + \frac{\delta_{\mathrm{B},n} + \delta_{\mathrm{R},n}}{2} \right]
\left[ 1 + a' (\delta_{\mathrm{B},n} - \delta_{\mathrm{R},n}) \right].
\end{equation}
Here, $a' = a/2 - b/a$ for the normalized discriminator and $a' = -b/a$ for the unnormalized one.

\begin{figure}[t]
\centering
\includegraphics[width=1\columnwidth]{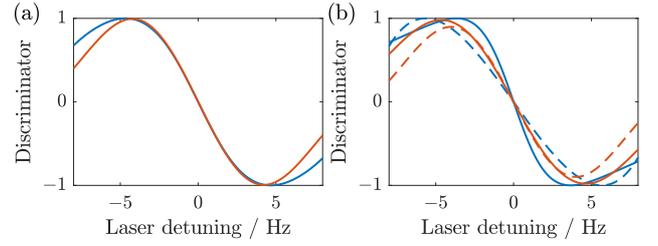}
\caption{Normalized (blue) and unnormalized (red) discriminators using the lineshape in Figure~\ref{fig:lineshape}: (a) For $\delta_{\mathrm{B},n} = \delta_{\mathrm{R},n} = 0$ and (b) for $(\delta_{\mathrm{B},n} - \delta_{\mathrm{R},n})/2\pi = 2$\;Hz (solid) and $(\delta_{\mathrm{B},n} - \delta_{\mathrm{R},n})/2\pi = -2$\;Hz (dashed).
\label{fig:discriminators}}
\end{figure}

Figure~\ref{fig:discriminators} plots the two discriminators \eqref{eq:Dis} using the lineshape in Figure~\ref{fig:lineshape} (not the series expansions). For $\delta_{\mathrm{B},n} = \delta_{\mathrm{R},n} = 0$, (a), they look very similar.
However, a laser-frequency flicker floor of $8\times 10^{-16}$ gives rise to differential noise $(\delta_{\mathrm{B},n} - \delta_{\mathrm{R},n})/2\pi$ beyond $\pm 2$\;Hz. This noise significantly changes the slope of the normalized discriminator, while it mainly reduces the amplitude of the unnormalized one, see Figure~\ref{fig:discriminators}(b).

A particular gain $G_\mathrm{o} = a^{-1}$ ($\approx 0.330\;\mathrm{FWHM}$ (full-width at half-maximum) for the ideal Rabi lineshape) corrects the offset between the laser and the transition frequency estimated by the discriminator  in a single step in the linear regime of the discriminator. We therefore define the fractional gain as $g = G\,a$. 

In many clocks, the clock frequency of cycle $n$ is considered to be the frequency where the transition was probed, which we will refer to as the `real-time' frequency,
$\omega_{\mathrm{rt},n} = \bar{\omega}_{\mathrm{l},n} + \omega_{\mathrm{s},n-1}$. Here the bar indicates that the free-running laser frequency is averaged over the cycle.
Another approach is to interpret the error term $GD_{i,n}$ not only as a servo parameter, but as an estimate of the offset between the transition and probing frequencies, which is added to the probing frequency at the end of the cycle to form the `post-processed' clock frequency, $\omega_{\mathrm{pp},n} = \omega_{\mathrm{rt},n} + GD_{i,n} = \bar{\omega}_{\mathrm{l},n} + \omega_{\mathrm{s},n}$. This approach is typically used with a fractional gain $g\approx 1$ and with a relatively high number of pulses per side \cite{Dube2015a}.
The servo errors of the real-time and post-processed frequencies can be expressed as $\Delta\omega_{\mathrm{rt},n} = \omega_{\mathrm{rt},n} - \omega_0 = \delta_{n-1} + (\bar{\delta}_{\mathrm{B},n} + \bar{\delta}_{\mathrm{R},n})/2$ and $\Delta\omega_{\mathrm{pp},n} = \omega_{\mathrm{pp},n} - \omega_0 = \delta_n + (\bar{\delta}_{\mathrm{B},n} + \bar{\delta}_{\mathrm{R},n})/2$, respectively. These differ in their response to linear drift, see \ref{sec:drift}, and in how they sample laser noise, see Figure~\ref{fig:noiseADEV}, but the noise-induced mean servo error is the same for both.

\section{Noise-induced servo error \label{sec:se}}

Next, we consider the servo error caused by laser noise with a zero mean.
In this case, the mean servo error will be equal to the mean of $\delta_n$.
For notational convenience, we define the mean and differential laser noise during servo cycle $n$ as $\bar{\delta}_n = (\bar{\delta}_{\mathrm{B},n} + \bar{\delta}_{\mathrm{R},n})/2$ and $\Delta_n = \bar{\delta}_{\mathrm{B},n} - \bar{\delta}_{\mathrm{R},n}$, respectively. For the normalized discriminator \eqref{eq:Dns}, the recursion formula for $\delta_n$ can then be written as
\begin{eqnarray}
\delta_n = & \left\{ 1-g \left[ 1 + \left( \frac{a}{2} - \frac{b}{a} \right) \Delta_{n} \right]\right\} \delta_{n-1} \nonumber \\
& - g \left(1+\frac{a}{2} \Delta_n\right) \bar{\delta}_{n} + g \frac{b}{a} \left( \frac{\widebar{\delta_{\mathrm{B},n}^2}}{2} - \frac{\widebar{\delta_{\mathrm{R},n}^2}}{2} \right).
\end{eqnarray}
To calculate the mean servo error, we recursively express the systematic detuning at cycle $N$ as
\begin{eqnarray}
\delta_N = & \prod_{i=1}^N \left\{ 1-g \left[ 1 + \left( \frac{a}{2} - \frac{b}{a} \right) \Delta_{i} \right]\right\} \delta_{0}  \nonumber \\
& - g \sum_{k=1}^{N-1} \prod_{i=k+1}^N \left\{ 1-g \left[ 1 + \left( \frac{a}{2} - \frac{b}{a} \right) \Delta_{i} \right]\right\} \nonumber \\
& \quad \quad \times \left[ \left(1+\frac{a}{2} \Delta_k\right) \bar{\delta}_{k} - \frac{b}{a} \left( \frac{\widebar{\delta_{\mathrm{B},k}^2}}{2} - \frac{\widebar{\delta_{\mathrm{R},k}^2}}{2} \right) \right] \nonumber \\
& - g \left[ \left(1+\frac{a}{2} \Delta_N\right) \bar{\delta}_{N} - \frac{b}{a} \left( \frac{\widebar{\delta_{\mathrm{B},N}^2}}{2} - \frac{\widebar{\delta_{\mathrm{R},N}^2}}{2} \right) \right]. \label{eq:recursion}
\end{eqnarray}
The first term vanishes in the limit $N \rightarrow \infty$ (with the time constant of the servo, $\tau_\mathrm{s} = -T_\mathrm{s}/\ln{(1-g)}$, where $T_\mathrm{s}$ is the servo cycle time) and thus we can assume a zero servo-initialization error,  $\delta_{0} = 0$, without loss of generality. We keep terms to first order in $a$ and $b/a$ only and  take the statistical average (over the servo index), where $\langle \bar{\delta}_{\mathrm{B},i}\rangle = \langle \bar{\delta}_{\mathrm{R},i}\rangle = 0$ and $\langle \widebar{\delta_{\mathrm{B},i}^2} \rangle = \langle \widebar{\delta_{\mathrm{R},i}^2} \rangle$. Finally, taking the limit $N \rightarrow \infty$, \eqref{eq:recursion} gives, after some manipulation, the mean servo error
\begin{eqnarray}
\langle \delta_\mathrm{s}\rangle = & - \frac{a}{4} \left( \langle \bar{\delta}_{\mathrm{B},i}^2\rangle - \langle \bar{\delta}_{\mathrm{R},i}^2\rangle \right) \nonumber \\
& + \frac{g}{2} \left( \frac{a}{2} - \frac{b}{a} \right) \sum_{m=1}^\infty (1-g)^{m-1} f_{m},  \label{eq:se}
\end{eqnarray}
where
\begin{eqnarray}
f_{m} = & \langle \bar{\delta}_{\mathrm{B},i} \bar{\delta}_{\mathrm{B},i+m}\rangle
+ \langle \bar{\delta}_{\mathrm{R},i} \bar{\delta}_{\mathrm{B},i+m}\rangle \nonumber \\
& - \langle \bar{\delta}_{\mathrm{B},i} \bar{\delta}_{\mathrm{R},i+m}\rangle
- \langle \bar{\delta}_{\mathrm{R},i} \bar{\delta}_{\mathrm{R},i+m}\rangle. \label{eq:fDm}
\end{eqnarray}
The first term in \eqref{eq:se}, which can also be written as $(-a/4) f_0$, is due to correlations between the noise during the B and R interrogations of the same cycle, while the second term is due to correlations between one cycle and all the subsequent ones. For $g=1$, only the first term in the sum over $m$ is nonzero, but also for smaller gains, the sum converges after only a few terms. Equation~\eqref{eq:se} can be applied to the unnormalized discriminator by omitting the terms proportional to $a$.

The servo error can thus be evaluated using the correlation function of the laser frequency noise $\Psi(k) = \langle \delta_\mathrm{l}(i\tau_0) \delta_\mathrm{l}((i+k)\tau_0)\rangle$, where $\delta_\mathrm{l}$ is the laser frequency noise and $\tau_0$ is the integration time. The mapping between the correlation functions in (\ref{eq:se}--\ref{eq:fDm}) and the lag index $k$ depends on the interrogation sequence. For example, for RB probing, the first term in \eqref{eq:se} is zero and for a clock where $n_\mathrm{t}$ transitions are interrogated, \eqref{eq:fDm} becomes
\begin{eqnarray} 
f_{m,\mathrm{RB},n_\mathrm{t}} =  & \Psi(2 n_\mathrm{t} m) + \Psi(2 n_\mathrm{t} m+1)
    \nonumber \\
&  - \Psi(2 n_\mathrm{t} m-1) - \Psi(2 n_\mathrm{t} m) .
\end{eqnarray}
The servo error is thus
\begin{eqnarray}
\langle \delta_{\mathrm{s},\mathrm{RB},n_\mathrm{t}}\rangle = \frac{g}{2} & \left( \frac{a}{2} - \frac{b}{a} \right)
 \sum_{m=1}^\infty (1-g)^{m-1} \nonumber \\
& \times \left[  \Psi(2 n_\mathrm{t} m+1) - \Psi(2 n_\mathrm{t} m-1)\right]. \label{eq:seRB}
\end{eqnarray}
On the other hand, for RBBR probing, the servo error becomes
\begin{eqnarray}
\langle \delta_{\mathrm{s},\mathrm{RBBR},n_\mathrm{t}}\rangle & =  - \frac{a}{8}
\left[ \Psi(1) -\Psi(3) \right] \nonumber \\
 & + \frac{g}{2} \left( \frac{a}{2} - \frac{b}{a} \right) \sum_{m=1}^\infty (1-g)^{m-1} f_{\Delta m,\mathrm{RBBR},n_\mathrm{t}}, \label{eq:seRBBR}
\end{eqnarray}
where
\begin{eqnarray}
f_{m,\mathrm{RBBR},n_\mathrm{t}} =
\frac{1}{4} [ & \Psi(4 n_\mathrm{t} m+1) + \Psi(4 n_\mathrm{t} m-1) \nonumber \\
& - \Psi(4 n_\mathrm{t} m+3) - \Psi(4 n_\mathrm{t} m -3) ]. \label{eq:fDmRBBR}
\end{eqnarray}
The integration time $\tau_0$ is the time spent interrogating at a particular frequency (see Section~\ref{sec:deadtime} for a justification). For a fixed total number of pulses per side, we therefore have $\tau_{0,\mathrm{RB}} = 2 \tau_{0,\mathrm{RBBR}}$. The correlation function as a function of lag index, $\Psi(k)$, depends on the integration time, so RB and RBBR probing must be compared using the relation $\Psi(k_\mathrm{RBBR} \tau_{0,\mathrm{RBBR}}) \approx \Psi(2 k_\mathrm{RB} \tau_{0,\mathrm{RBBR}})$, where the equality is approximative due to the different cut-off frequencies, see Section~\ref{sec:corrfunc}.

For the higher-order Thue-Morse sequence RBBRBRRB, used, e.g., in \cite{Baynham2019PhD}, the first term in \eqref{eq:se} is zero and \eqref{eq:fDm} becomes
\begin{eqnarray}
f_{m,\mathrm{RBBRBRRB},n_\mathrm{t}} = \frac{1}{16}
[ \Psi(8 n_\mathrm{t} m+1) - \Psi(8 n_\mathrm{t} m-1) \nonumber \\
\quad \quad + \Psi(8 n_\mathrm{t} m-3) - \Psi(8 n_\mathrm{t} m+3) + \Psi(8 n_\mathrm{t} m-5) \nonumber \\
\quad \quad - \Psi(8 n_\mathrm{t} m+5) + \Psi(8 n_\mathrm{t} m+7) - \Psi(4 n_\mathrm{t} m -7) ]. \label{eq:fDmRBBRBRRB}
\end{eqnarray}
More complex probing sequences can be treated similarly. If the transitions are interrogated with different pulse lengths, see, e.g., \cite{Steinel2022a}, the noise integration time $\tau_0$ must be chosen such that the time spent at each interrogation frequency is an (approximate) multiple of it.

From \eqref{eq:se}, a significantly larger servo error is expected when the normalized discriminator is used due to the terms proportional to $a$, which are significantly larger than the $b/a$ term and are only present when the normalized discriminant is used. Also, for RBBR probing with the normalized discriminator, the first term in \eqref{eq:seRBBR}, which is independent of the gain $g$, can be very significant. As the time between consecutive servo updates is shorter when a single transition is probed, the correlations in the $f_{\Delta m}$ term are stronger and the servo error will be larger. This is the opposite to the servo error caused by a linear drift, where the error is (approximately) proportional to the cycle time, see~(\ref{eq:drift-rt}--\ref{eq:drift-pp}).

The normalized discriminator is used to prevent fluctuations in the excitation probability from changing the effective servo gain. The servo error can be reduced while retaining insensitivity to slow fluctuations by normalizing by a moving mean of the $M$ latest transition probabilities. In this case, the $a/2$ term in \eqref{eq:Dns} becomes $a/(2M) \sum_{i = 0}^{M-1} \Delta_{n-i}$, and the servo error can be evaluated to be
\begin{eqnarray}
\langle \delta_\mathrm{s}\rangle = & - \frac{a}{4} \frac{1}{M} \sum_{i=0}^{M-1} f_{-i} \nonumber \\
& + \frac{g}{2} \sum_{m=1}^\infty (1-g)^{m-1} \left( \frac{a}{2} \frac{1}{M} \sum_{i=0}^{M-1} f_{m-i} - \frac{b}{a} f_{m} \right).  \label{eq:se-mm}
\end{eqnarray}
This shows how the terms proportional to $a$ are reduced by distributing the correlation functions over a range of lag indices $k$.

An intuitive way to cancel the error due to correlations (for both discriminators) is to reverse the interrogation order on every second servo cycle, e.g., from RB to BR or RBBR to BRRB (note that alternating RB/BR probing is not the same as RBBR probing, as there is an additional servo update between RB and BR).
This can be analyzed using a recursion similar to \eqref{eq:recursion}, but with the sign of the differential noise reversed for even cycles. The mean error is then obtained as the mean of $\langle \delta_N \rangle$ and $\langle \delta_{N+1} \rangle$ in the limit $N \rightarrow \infty$, which is identically zero to this order of expansion.

\section{Noise \label{sec:noise}}

\subsection{Laser noise} \label{sec:laser-noise}

The fundamental instability limit of an ultrastable laser stabilized to a Fabry--Perot cavity is due to the cavity length fluctuations caused by thermal Brownian noise~\cite{Numata2004a}, which has a flicker-frequency behaviour. Cavities are typically made of ultra-low expansion glass (ULE) spacers of $100\ldots 300$\;mm length, although longer ULE cavities~\cite{Hafner2015a} as well as cryogenic silicon cavities \cite{Kessler2012a} have been demonstrated. For a 100-mm ULE cavity, the thermal-noise-limited instability is approximately $4\times 10^{-16}$. A possible white frequency noise contribution is uncorrelated and will not contribute to the servo error, so we consider flicker frequency noise only.

ULE cavities exhibit isothermal drift, typically on the order of $10\;\mathrm{mHz/s}$, due to the amorphous nature of the material. This drift can be compensated using a second-order integrating servo~\cite{Peik2006a} or by evaluating the drift against the atoms/ion using an exponentially weighted moving average~\cite{Dube2017a} and will not be considered here. The drift-compensation will, however, lead to high-pass filtering of the laser noise.

\subsection{Correlation function of flicker noise} \label{sec:corrfunc}

Flicker ($1/f$) noise of infinite bandwidth is a non-stationary process with a diverging variance. A low cut-off frequency is thus needed. As explained in \cite{Vernotte2015b}, removing the mean value of a sequence of flicker noise is equivalent to setting the amplitude at $f=0$ to zero or to setting the low cut-off frequency to the inverse of the duration of the sequence. In the case of laser noise, the drift compensation mentioned in Section~\ref{sec:laser-noise} will set the low cut-off frequency $f_\mathrm{l}$ to the inverse of the drift-compensation time constant. This has the benefit that the correlation function is independent of the duration of the noise sequence. A high cut-off frequency of $f_\mathrm{h} = 1/(2 \tau_0)$ is imposed by the integration time $\tau_0$ \cite{Vernotte2015b}.

For the high-pass--filtered flicker noise, we write the fractional-frequency power spectral density (PSD) as
\begin{equation} \label{eq:Sy}
S_y(f) = \left\{
    \begin{array}{ll}
    0 & \mathrm{for}\; f < s f_\mathrm{l}, \\
    h_{-1} \frac{f-s f_\mathrm{l}}{(1-s) f_\mathrm{l}^2} & \mathrm{for}\; s f_\mathrm{l} < f < f_\mathrm{l}, \\
    h_{-1}/f & \mathrm{for}\; f_\mathrm{l} < f < f_\mathrm{h}, \\
    0 & \mathrm{for}\; f > f_\mathrm{h},
    \end{array}
    \right.
\end{equation}
where the filter is approximated to have a linear transition with steepness $s$ ($s=0.5$ for all numerical results) and the PSD coefficient $h_{-1}$ corresponds to a flicker floor of $[2 \ln{(2)} \, h_{-1}]^{1/2}$ in the Allan deviation.
The correlation function is the Fourier transform of the PSD, $\Psi_y(\tau) = \int_0^\infty S_y(f) \cos{(2 \pi\tau f)} \mathrm{d}f$, and becomes, following \cite{Vernotte2015b},
\begin{eqnarray}
\Psi_y(\tau) = h_{-1} & \bigg[ \frac{\cos{(\omega_\mathrm{l}\tau)} - \cos{(s \omega_\mathrm{l}\tau)} +
    \omega_\mathrm{l}\tau (1-s) \sin{(\omega_\mathrm{l}\tau})}{(1-s)(\omega_\mathrm{l}\tau)^2} \nonumber \\
    & \; + \Ci{(\omega_\mathrm{h}\tau)} - \Ci{(\omega_\mathrm{l}\tau)} \bigg], \label{eq:Psi} \\
\Psi_y(0) = h_{-1} & \left[ \frac{1-s}{2} + \ln{\frac{\omega_\mathrm{h}}{\omega_\mathrm{l}}} \right], \label{eq:Psi0}
\end{eqnarray}
where $\omega_\mathrm{l,h} = 2\pi f_\mathrm{l,h}$ and the cosine integral function $\Ci{(x)}$ is defined as
\begin{equation}
\Ci{(x)} = -\int_x^\infty \frac{\cos{y}}{y} \mathrm{d}y, \quad x>0.
\end{equation}
If $f_\mathrm{l}$ is very low, \eqref{eq:Psi} can be approximated as
\begin{equation} \label{eq:Psi_appr}
\Psi_y(\tau) \approx h_{-1} \left[ \frac{1-s}{2} -\gamma - \ln{|\omega_\mathrm{l}\tau|} \right],
\end{equation}
where $\gamma \approx 0.5772$ is the Euler--Mascheroni constant \cite{Vernotte2015b}.

\begin{figure}[tb]
\centering
\includegraphics[width=0.95\columnwidth]{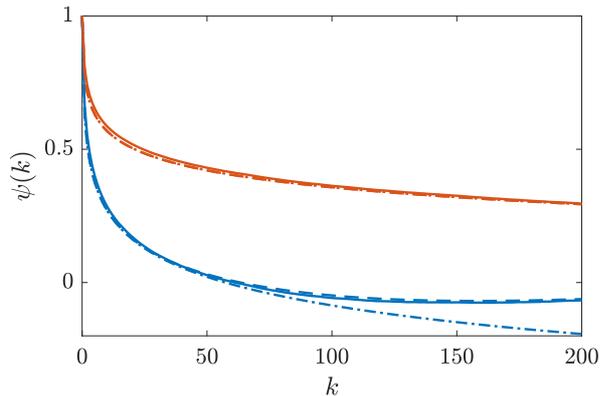}
\caption{Normalized correlation functions for $\tau_0 = 2$\;s and $f_\mathrm{l} = (1000\;\mathrm{s})^{-1}$ (blue) and $f_\mathrm{l} = (1\;\mathrm{d})^{-1}$ (red). The solid lines are calculated numerically from 100~days of simulated data, the dashed lines are from (\ref{eq:Psi}--\ref{eq:Psi0}), and the dash-dotted lines are from the approximation \eqref{eq:Psi_appr}. The red dashed and dash-dotted lines overlap. 
\label{fig:corrfunc}}
\end{figure}

To express the correlation function as a function of the lag index $k=\tau/\tau_0$, one only has to substitute $\omega_\mathrm{h}\tau = k\pi$ and $\omega_\mathrm{l}\tau = 2 k \pi/N_\mathrm{l}$, where $N_\mathrm{l} = (f_\mathrm{l} \tau_0)^{-1}$.
Figure~\ref{fig:corrfunc} shows normalized correlation functions, $\psi(k) = \Psi_y(k)/\Psi_y(0)$, for flicker with $\tau_0 = 2$\;s and with two vastly different low cut-off frequencies corresponding to time constants of 1000\;s and 1\;d. Numerically calculated correlation coefficients require very long simulated data sets\footnote{All simulated noise was generated using the MATLAB function \texttt{f\_alpha\_gaussian.m}  \cite{Stoyanov2011a} based on \cite{Kasdin1995a}.} for good accuracy (here 100 days), but agree quite well with the analytical expressions (\ref{eq:Psi}--\ref{eq:Psi0}).
The difference compared to the analytical result changes if one varies the filter steepness $s$, indicating that it is due to the simplistic filter model rather than the quality of the simulated noise.
The approximation \eqref{eq:Psi_appr} fails for $k>50$ for the short time constant ($f_\mathrm{h}/f_\mathrm{l} = 250$), but is good up to $k=500$ for $f_\mathrm{h}/f_\mathrm{l} \gtrsim 5000$ (the 1-d time constant gives $f_\mathrm{h}/f_\mathrm{l} = 21\,600$).
Since the sum in the servo error \eqref{eq:se} converges quickly, \eqref{eq:Psi_appr} will give a reasonable approximation in many cases. As the servo error consists of differences between correlation functions evaluated at different times, the dependence on $\omega_\mathrm{l}$ cancels, $\Psi_y(k)-\Psi_y(k') \approx h_{-1} \ln{|k'/k|}$, and the servo error will depend only on the lineshape parameters $a$ and $b$, the flicker $h_{-1}$, the gain $g$ and the interrogation sequence. However, for this work, \eqref{eq:Psi_appr} offers no significant numerical advantage over the exact expressions.

\subsection{Dead time} \label{sec:deadtime}

\begin{figure}[tb]
\centering
\includegraphics[width=0.95\columnwidth]{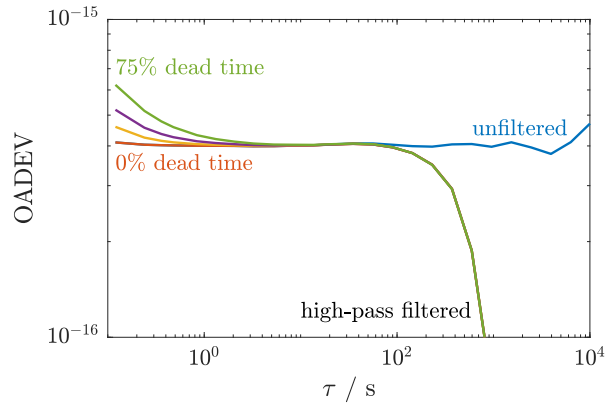}
\caption{Overlapping Allan deviation of the unfiltered simulated flicker frequency noise and the high-pass--filtered noise with relative dead times of 0\%, 25\%, 50\%, and 75\%.
\label{fig:deadtime}}
\end{figure}

Optical clocks typically have dead time, required for state detection, cooling and state preparation, between the interrogation periods. We must therefore investigate how this affects the laser noise statistics. To do this, we simulate flicker frequency noise on a 30-ms grid and assume that each interrogation lasts 120\;ms, including a relative dead time of 0\%, 25\%, 50\%, or 75\%. The laser noise is then resampled to the 120-ms grid by averaging the samples belonging to the probe pulse (4, 3, 2, or 1 out of 4). Figure~\ref{fig:deadtime} shows the overlapping Allan deviation (OADEV) of the unfiltered laser noise and the high-pass--filtered (with 1000-s time constant) noise with different dead times. We see that the dead time gives rise to a small white-frequency-noise contribution. As white noise is uncorrelated, it will not contribute to the servo error, which was verified by servo simulations.  We also verified that the impact on the correlation function is insignificant.  Thus, we can neglect the dead time and use the time spent interrogating at a certain frequency (i.e., $n_\mathrm{p}$ pulses on the blue or red side of a particular transition) as the integration time $\tau_0$,
which significantly speeds up the numerical simulations.

\subsection{Magnetic-field noise}

For optical clock transitions with linear Zeeman shift, the magnetic field strength is continuously measured through the frequency splitting of one or more symmetric pairs of transitions. In the VTT $^{88}$Sr$^+$ clock \cite{Lindvall2022a}, the magnetic-field noise is well described by an 11-pT flicker floor and small  temperature-correlated fluctuations. Similar behaviour, although with higher noise levels, has been observed elsewhere \cite{Barwood2015a}. Assuming the noise to be flicker dominated, we can apply the results of Section~\ref{sec:corrfunc} directly by multiplying the magnetic-field noise by the Zeeman sensitivity of the transition to convert it to frequency noise. If the different Zeeman components have separate second-order integrating servos, the magnetic-field noise is high-pass--filtered like the laser noise. Otherwise, the length of the data set determines the low cut-off frequency, but, as will be discussed in Sec.~\ref{sec:results}, the choice of cut-off frequency is typically not critical.

If, in a particular clock, (high-pass--filtered) random-walk magnetic-field noise is significant, it can be analyzed using the same methods used for flicker noise here.

\section{Servo simulations} \label{sec:sim}

To support the findings of the analytical servo error \eqref{eq:se} and to test the limitations of the involved approximations, servo simulations were carried out. These can take as input deterministic laser-frequency and magnetic-field changes as well as noise and then simulate the operation of the optical clock. QPN can be chosen to be included or excluded. The simulations were carried out for a probe pulse length of 100\;ms, a dead time of 25\;ms, the ideal Rabi lineshape of Figure~\ref{fig:lineshape} and with other parameters as mentioned in Section~\ref{sec:results}. The length of the simulations was typically 48\;h and $100$ simulations were averaged in each case.
The simulation results are presented in Section~\ref{sec:results} compared to the results from the analytical expressions.
To minimize the uncertainty of the servo error, these do not include QPN, but it was separately verified that simulations including QPN do agree with those without within the combined uncertainty.

Simulations were also carried out with unfiltered laser noise and a second-order servo for drift compensation. With the same drift-compensation time constant, this gave essentially identical servo errors as high-pass--filtered noise and no second-order servo, which verifies that high-pass--filtering is an adequate and simple model for the drift compensation that enables the analytical servo-error analysis.

\section{Numerical results} \label{sec:results}

The servo error \eqref{eq:se} is applicable to any optical clock using Rabi interrogation when it comes to laser noise and to clocks with linear Zeeman shift when it comes to magnetic-field noise. However, few publications contain sufficient details on the interrogation sequence and noise properties to evaluate the noise-induced shifts.

\begin{figure}[tb]
\centering
\includegraphics[width=0.95\columnwidth]{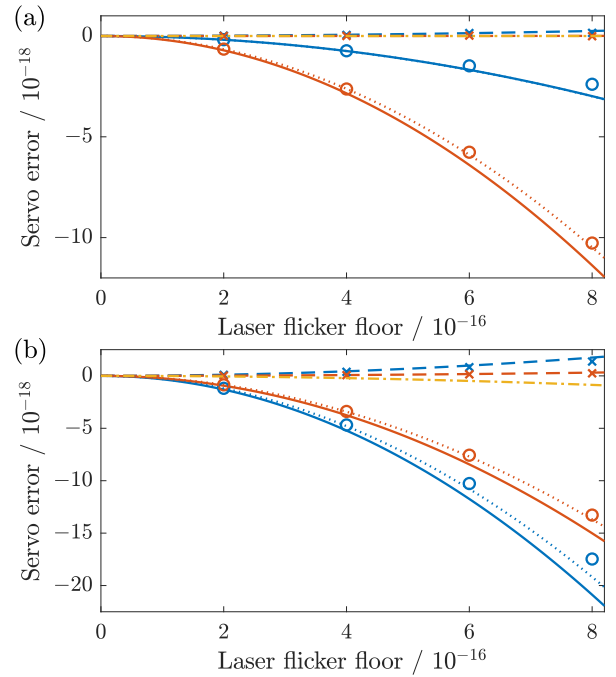}
\caption{Servo error for a clock based on (a) six Zeeman components and (b) a single transition as a function of the laser flicker floor with the gain $g=1$. Blue (red) lines/symbols correspond to RB (RBBR) probing. Solid (dashed) lines are analytical results (\ref{eq:seRB}--\ref{eq:fDmRBBR}) for the normalized (unnormalized) discriminator; circles (crosses) are from the corresponding servo simulations (mean of 100 two-day simulations). Dotted lines are the analytical formulas for the normalized discriminator evaluated using numerical correlation functions (in (a), solid and dotted blue lines overlap).
For RBBRBRRB probing, only the analytical result \eqref{eq:fDmRBBRBRRB} for the normalized discriminator is shown (dash-dotted yellow line; overlaps with dashed red line in (a)).
\label{fig:laser-flicker}}
\end{figure}

\begin{figure}[tb]
\centering
\includegraphics[width=0.95\columnwidth]{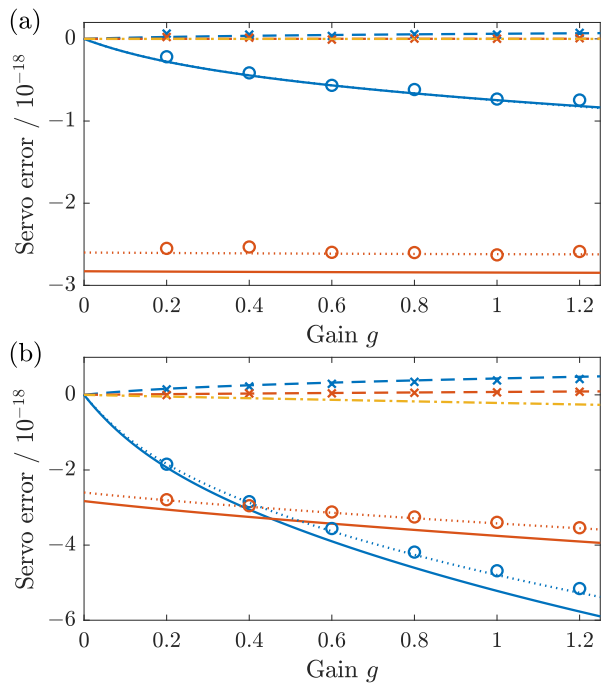}
\caption{Servo error for a clock based on (a) six Zeeman components and (b) a single transition as a function of the gain when the laser flicker floor is $4\times 10^{-16}$.
Blue (red) lines/symbols correspond to RB (RBBR) probing. Solid (dashed) lines are analytical results for the normalized (unnormalized) discriminator; circles (crosses) are from the corresponding servo simulations.
Dotted lines are the analytical formula for the normalized discriminator evaluated using numerical correlation functions (in (a), solid and dotted blue lines overlap).
For RBBRBRRB probing, only the analytical result for the normalized discriminator is shown (dash-dotted yellow line; overlaps with dashed red line in (a)).
\label{fig:gain}}
\end{figure}

\begin{figure}[tb]
\centering
\includegraphics[width=0.95\columnwidth]{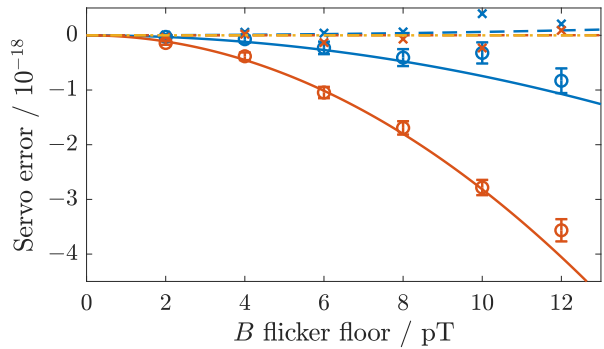}
\caption{Mean servo error for a $^{88}$Sr$^+$  clock based on six Zeeman components as a function of the magnetic-field flicker floor for $g=1$. Blue (red) lines/symbols correspond to RB (RBBR) probing. Solid (dashed) lines are analytical results for the normalized (unnormalized) discriminator; circles (crosses) are from the corresponding servo simulations.
For RBBRBRRB probing, only the analytical result for the normalized discriminator is shown (dash-dotted yellow line; overlaps with dashed red line).
\label{fig:B-noise}}
\end{figure}

The numerical results will be presented for a $^{88}$Sr$^+$ single-ion clock, studied at the National Research Council (Canada) \cite{Dube2015a,Jian2023a}, the National Physical Laboratory (UK) \cite{Barwood2014a,Barwood2015a}, the Physikalisch-Technische Bundesanstalt (Germany) \cite{Steinel2022a}, and by us at VTT \cite{Lindvall2022a}. These results are directly applicable also to $^{40}$Ca$^+$ and $^{138}$Ba$^+$ clocks using Rabi interrogation. We consider a clock, where the $S_{1/2} \rightarrow D_{5/2}$ transitions
$m_J = \mp 1/2 \rightarrow m_J = \mp 1/2$, $\pm 1/2 \rightarrow \pm 3/2$, and $\pm 1/2 \rightarrow \pm 5/2$ with Zeeman sensitivities $\pm 5.6$, $\pm 11$, and $\pm 28\;\mathrm{kHz/\mu T}$ are interrogated in order to cancel the linear Zeeman shift as well as the electric quadrupole shift and other tensor shifts.
For simplicity, we use an interrogation sequence where the six Zeeman components are probed with the same pulse length and the same number of times during a servo cycle, and their frequencies are averaged to obtain the unperturbed clock line center. The individual components are either probed as RB, RBBR, or RBBRBRRB and the two discriminators (\ref{eq:Dn}--\ref{eq:Du}) are compared. For RBBRBRRB probing, the servo error is small and for clarity, only the analytical result for the normalized discriminator is plotted in the figures, but the simulation results were found to be in good agreement.
The number of interrogation pulses per frequency vary between groups.
In Figures \ref{fig:laser-flicker}--\ref{fig:B-noise}, the number of pulses per side is 16 (split into 2 sets of 8 for RBBR probing and 4 sets of 2 for RBBRBRRB).

For comparison, we also show the corresponding results for a clock based on a single transition using the same parameters, whose cycle time then is shorter by a factor of 6. This could be, e.g., a clock based on the $^{171}$Yb$^+$ electric octupole (E3) transition interrogated using Rabi spectroscopy \cite{King2012a}.
Clocks where a single pair of Zeeman components are interrogated, e.g., the $^{27}$Al$^+$ and  $^{115}$In$^+$ ion clocks and fermionic optical lattice clocks,
would fall between the single- and six-transition clocks.

The servo errors are given as fractional frequencies relative to the 445-THz $^{88}$Sr$^+$ clock frequency. As the error is proportional to the square of the noise, it will for other species scale linearly with the clock frequency for a constant fractional laser-frequency flicker floor. E.g., for the 642-THz $^{171}$Yb$^+$ E3 transition, the fractional error would be a factor of 1.4 as large. In the case of magnetic-field noise, the error scales with the square of the Zeeman sensitivity.

We limit ourselves to noise levels that have little effect on the linewidth so that the lineshape expansion parameters (HWHM, $p_0$, $a$, $b$) can be kept constant.
A laser flicker floor of $8\times 10^{-16}$ at 445\;THz corresponds to a Gaussian lineshape with FWHM ${\sim} 1$\;Hz for a 10-s integration time~\cite{Domenico2010a}, which has a negligible effect on the slopes of the lineshape in Figure~\ref{fig:lineshape}.
The analytical expression for the servo error \eqref{eq:se} is expected to give quantitative results when the long-term noise distribution lies within the region where the series expansion describes the lineshape well. This corresponds to noise levels slightly lower than the lineshape criterion above.

\begin{figure}[tb]
\centering
\includegraphics[width=1\columnwidth]{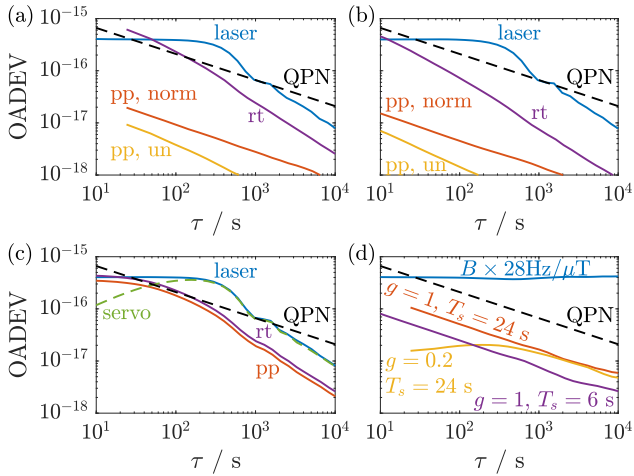}
\caption{(a--c) Overlapping Allan deviation of the high-pass--filtered laser frequency flicker ($4\times 10^{-16})$ and the different servo errors (rt---real-time, pp---post-processed, norm/un---normalized/unnormalized discriminator) compared to the QPN limit for (a) $g=1$ and $T_\mathrm{s}=24$\;s, (b) $g=1$ and $T_\mathrm{s}=6$\;s, (c) $g=0.2$ and $T_\mathrm{s}=6$\;s.
Both discriminators give practically the same OADEV for the real-time frequency and in (c) also for the post-processed one.
In (c), the OADEV of the line-center servo frequency, i.e., the mean of $\omega_{\mathrm{s},n}$ for the six transitions, is shown (dashed green curve); in (a--b), it would overlap with the laser curve. (d) For 6.4-pT magnetic-field flicker, all servo errors have essentially identical instabilities and a single curve is plotted for each set of servo parameters as indicated in the figure. (b) and (d) share the vertical-axis labels with (a) and (b), respectively.
\label{fig:noiseADEV}}
\end{figure}

Before studying the magnitude of the servo errors, Figures~\ref{fig:noiseADEV}(a--c) demonstrate how the errors due to laser frequency flicker average down towards their mean values for the six-component $^{88}$Sr$^+$ clock. For $g=1$, the laser noise is well cancelled from the post-processed frequency (with the unnormalized discriminator performing better), whereas the real-time frequency samples the laser noise differentially (the ion samples it one cycle late). This can degrade the short-term stability of the real-time frequency for long servo cycle times, see Figure~\ref{fig:noiseADEV}(a), where the real-time curve is above the QPN limit for short $\tau$. On the other hand, the real-time servo error averages down approximately as $\tau^{-1}$ due to the differential sampling. Shortening the cycle time reduces the sampled noise as shown in Figure~\ref{fig:noiseADEV}(b). For a small gain $g=0.2$, the error term does not apply the full correction and the instabilitites of the real-time and post-processed servos are very similar, see Figure~\ref{fig:noiseADEV}(c). For simulations of $6\ldots 48$\;h duration (about $1000\ldots 7000$ cycles with $T_\mathrm{s}=24$\;s), the run-to-run variations of the mean servo error are similar for both the real-time and post-processed frequencies and for both discriminators, and a factor ${\lesssim} 5$ smaller than the corresponding QPN limit for the noise levels and gains considered here.

For magnetic-field flicker noise, there are no significant differences between the two discriminators or between the real-time and post-processed frequencies. Figure~\ref{fig:noiseADEV}(d) shows the servo-error OADEV for different servo parameters and a magnetic-field flicker of 6.4\;pT, chosen to give the same frequency noise for the most sensitive interrogated Zeeman component as the laser noise in Figure~\ref{fig:noiseADEV}(a--c).
The run-to-run variations of the servo error are independent of the gain $g$ and larger than for a similar servo error caused by laser noise, but still below the QPN limit.
In the following, the simulated results are shown for the post-processed frequency.

Figure~\ref{fig:laser-flicker} shows the mean servo error as a function of the laser flicker floor for the different clocks, probing sequences, and discriminators when the gain is $g=1$.
The normalized discriminator and the single-transition clock are very susceptible to this servo error: for RB probing, the model predicts servo errors of $-5\times 10^{-18}$ and $-0.3\times 10^{-18}$ for laser noise floors of $4\times 10^{-16}$ and $1\times 10^{-16}$, respectively.
Figure~\ref{fig:laser-flicker} also shows that when the servo error is dominated by correlations within the same cycle (RBBR probing), the error is similar for the six- and single-transition clocks, but when it is dominated by the cycle-to-cycle correlations (RB, RBBRBRRB), the error is significantly larger for the single-transition clock, as anticipated in Section~\ref{sec:se}.
The drift-compensation time constant was here taken to be 960\;s in all cases.

The difference between the analytical and numerical results is partially caused by the difference between the simple high-pass--filter model in \eqref{eq:Sy} and the actual filter used for the simulated noise, as mentioned in the context of Figure~\ref{fig:corrfunc}.
This is illustrated by the dotted lines in Figure~\ref{fig:laser-flicker}, which correspond to the analytical formulas evaluated using numerically calculated correlation functions as in Figure~\ref{fig:corrfunc}. Accounting for this, the agreement between the analytical and numerical results is very good up to a laser-frequency flicker floor of ${\sim}6\times 10^{-16}$ as expected, while the analytical formula overestimates the error for higher noise levels.

In the regime where the approximative correlation function \eqref{eq:Psi_appr} is valid, the servo error is independent of the integration time $\tau_0$ and thus on the number of pulses per side. Using the exact correlation function \eqref{eq:Psi}, a small dependence on $\tau_0$ comes from the ratio of the high and low cut-off frequencies. For a 960-s drift-compensation time constant and between 4 and 32 pulses per side, the absolute change in the error is ${\lesssim} 2\times 10^{-20}$ for all sequences considered. This has also been verified by servo simulations, where the change in servo error is smaller than the statistical uncertainty.
However, for a fixed simulation duration, the run-to-run variations of the servo error are smaller for lower numbers of pulses (shorter cycle times), as expected from Figures~\ref{fig:noiseADEV}(a--b).

Figure~\ref{fig:gain} shows the mean servo error as a function of the gain $g$ for the same combinations as in Figure~\ref{fig:laser-flicker}. When the correlations between consecutive cycles dominate (RB, RBBRBRRB probing), the error decreases with decreasing gain, whereas the servo error is constant when the correlations between R and B noise dominate (RBBR probing of six components). Again, the agreement between model and simulations becomes excellent when the numerically evaluated correlation function is used (dotted lines).

The dependence on magnetic-field flicker is similar to that on laser flicker with three main differences. First, the servo error varies strongly for different Zeeman components due to the scaling with the square of the Zeeman sensitivity, and the mean error will thus be dominated by the most sensitive components interrogated.
Secondly,  the low cut-off frequency of the noise is lower unless the components are tracked by individual second-order servos. Finally, the variations of the error is larger for magnetic-field noise, as discussed in the context of Figure~\ref{fig:noiseADEV}.

The mean servo error for a $^{88}$Sr$^+$ clock is shown in Figure~\ref{fig:B-noise}. The highest noise floor of 13\;pT has the same effect on the most sensitive transitions as a laser frequency flicker floor of $8\times 10^{-16}$. Significantly higher noise levels have been observed \cite{Barwood2014a,Steinel2022a}, but then the interrogation time must be limited below the 100\;ms considered here, at least for the most sensitive transitions.
Here, no high-pass--filtering was used, so a low cut-off frequency corresponding to the simulation duration of 2\;d was used for the analytical results, but it turns out that the result is essentially independent of the time constant as long as this is ${\gtrsim} 500 \tau_0$. This is the regime where \eqref{eq:Psi_appr} is valid for all significant terms in \eqref{eq:se}. This also applies to the results for laser flicker, which makes the choice of cut-off frequency less critical.
For the simulations with the normalized discriminator, the uncertainty of the mean of 100 simulations is shown as error bars. Comparing to Figure~\ref{fig:laser-flicker}, the larger scatter of the simulated data points, as expected from Figure~\ref{fig:noiseADEV}, is evident.

Magnetic-field noise is particularly important for the alkaline-earth-metal ion clocks ($^{40}$Ca$^+$, $^{88}$Sr$^+$, $^{138}$Ba$^+$, and $^{226}$Ra$^+$) due to their large linear Zeeman shifts, and multiple layers of magnetic shielding is typically used in these. The recently demonstrated clock based on the highly charged Ar$^{13+}$ ion also has linear Zeeman shifts of similar magnitude~\cite{King2022a}. In other clocks where a pair of Zeeman components are interrogated, e.g., $^{27}$Al$^+$, $^{115}$In$^+$, and fermionic optical lattice clocks, the Zeeman sensitivities are 2--4 orders lower, but the magnetic-field noise can potentially be significantly higher since magnetic shielding is typically not used.

The alternating schemes RB/BR, RBBR/BRRB, and RBBRBRRB/BRRBRBBR were also simulated. In all cases, the alternating probing gave servo errors consistent with zero ($\chi^2 = 0.78$ for 106 points like those in the figures, each the average of 100 simulations) and with varying sign, indicating that the mean servo error is indeed zero. The run-to-run variations are, however, of similar magnitude as for the corresponding fixed probing schemes.

\begin{figure}[tb]
\centering
\includegraphics[width=0.95\columnwidth]{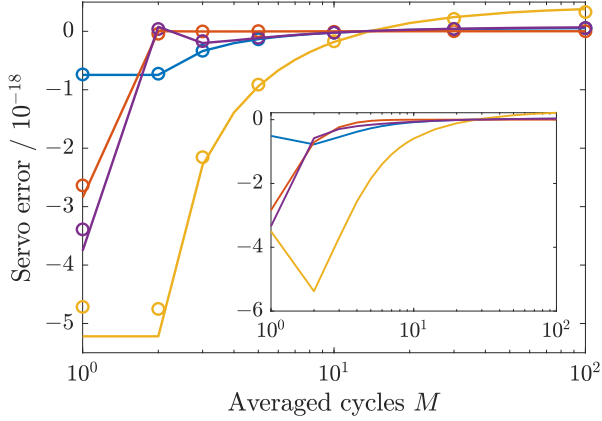}
\caption{Servo error of the moving-mean-normalized discriminator as a function of the number of averaged cycles $M$ for six-component RB (blue) and RBBR probing (red) as well as single-transition RB (yellow) and RBBR probing (purple). Circles are from servo simulations, lines from the analytical expression \eqref{eq:se-mm}.
The laser-frequency flicker floor is $4\times 10^{-16}$ and the gain $g=1$. The inset shows the corresponding analytical results for $g = 0.5$.
\label{fig:movmean}}
\end{figure}

Figure~\ref{fig:movmean} illustrates how moving-mean normalization decreases the servo error of the normalized discriminator for six- and single-transition clocks and for RB and RBBR probing. The most striking difference is that averaging two cycles does nothing for RB probing, while it cancels most of the error for RBBR probing. The other interesting observation is that all the curves are approximately zero for $M \sim 15$ for $g=1$. For smaller gains, the zero crossing occurs at higher $M$. For large $M$, all curves approach the corresponding values for the unnormalized discriminator.

\section{Discussion and summary \label{sec:summary}}

For simplicity, the numerical results were presented for an ideal Rabi lineshape without considering the atom/ion temperature or the natural lifetime of the clock state.
This is justified, as the largest servo errors are caused by the normalized discriminator, where the error is approximately proportional to the square of the slope and thus not very sensitive to the exact lineshape. For comparison, a brief treatment of Ramsey interrogation, for which the noise-induced servo error is found to be negligible, is included in \ref{app:Ramsey}.

As the noise-induced servo error depends on the details of the lineshape, the statistics of the noise, and the method of drift compensation, it is difficult to evaluate exactly. Hence, a scheme that cancels it is preferred. For this, the following approach can be used: (i) Use the unnormalized discriminator \eqref{eq:Du} or normalize by a moving mean of probabilities.
(ii) Tailor the interrogation sequence to be insensitive against the effects relevant for the particular clock, such as laser frequency drift, intensity drift, and magnetic-field drift. A Thue--Morse sequence of length 8 (RBBRBRRB) will also significantly reduce the noise-induced servo error. (iii) Alternate the probing order, e.g., between RB and BR, for every second servo cycle in order to cancel any remaining noise-induced error.

To summarize, we have studied servo errors caused by the correlations of laser-frequency and magnetic-field flicker noise in combination with an imbalance in the response of the Rabi-interrogation servo to positive and negative differential noise. We derived an analytical estimate for the servo error that can be applied to arbitrary lineshapes and interrogation sequences and to any noise statistics whose correlation function can be evaluated, and compared it to numerical servo simulations with good agreement.  We showed that the commonly used normalized servo discriminator significantly increases these errors, but that these can be canceled by normalizing using a moving mean or by alternating the probing order.

\section*{Data availability statement}

The data that support the findings of this study are openly available \cite{data}.

\ack

We thank Nils Huntemann for reading and commenting the manuscript.
This work was supported by the project 20FUN01 TSCAC, which has received funding from the EMPIR programme co-financed by the Participating States and from the European Union’s Horizon 2020 research and innovation programme.
It was also supported by the Academy of Finland (REASON, decision 339821) and is part of
the Academy of Finland Flagship Programme “Photonics Research and Innovation” (PREIN, decision 320168).

\appendix

\section{Linear laser drift \label{sec:drift}}

\setcounter{section}{1}

Here, we consider a linear laser drift $d$ and RB probing.
In \eqref{eq:Dis}, we use the noise terms to describe the effect of the drift: $(\delta_{\mathrm{B},n} + \delta_{\mathrm{R},n})/2 = n d T_\mathrm{s}$ is the mean laser frequency during the cycle $n$ interrogation of a particular transition (with an arbitrary offset that is compensated by $\delta_{n-1}$) and $\delta_{\mathrm{B},n} - \delta_{\mathrm{R},n} = d T_\mathrm{s}/(2 n_\mathrm{t})$ is the frequency change between the R and B interrogations, where $n_\mathrm{t}$ is the number of interrogated transitions.
In the steady state, the discriminator is constant, $D_{i,n+1}= D_{i,n}$, which leads to
\begin{equation}
G D_{i,n} = - g (\delta_{n-1} + n d T_\mathrm{s})\left(1+ a' \frac{d T_\mathrm{s}}{2 n_\mathrm{t}} \right) = -d T_\mathrm{s}.
\end{equation}
Solving for $\delta_{n-1}$ gives $\delta_{n-1} = d T_\mathrm{s} ( 1 - ng')/g'$, where the `drift-dependent gain' is $g' = g [ 1 + a' d T_\mathrm{s}/(2 n_\mathrm{t})]$. BR probing gives a similar expression with a minus sign for the $a'$ term.
The real-time and post-processed servo errors are then independent of $n$ as expected for a constant drift
\begin{eqnarray}
\Delta\omega_{\mathrm{rt}} &=& \delta_{n-1} + n dT_\mathrm{s} = d T_\mathrm{s} \frac{1}{g'} \label{eq:drift-rt}, \\
\Delta\omega_{\mathrm{pp}} &=& \delta_{n-1} + (n-1) dT_\mathrm{s} = d T_\mathrm{s} \frac{1 - g'}{g'} \label{eq:drift-pp}.
\end{eqnarray}
In \cite{Yuan2021a}, $\Delta\omega_{\mathrm{rt}}$ was considered , while $\Delta\omega_{\mathrm{pp}}$ was used in \cite{Dube2018a}.

\begin{figure}[tb]
\centering
\includegraphics[width=1\columnwidth]{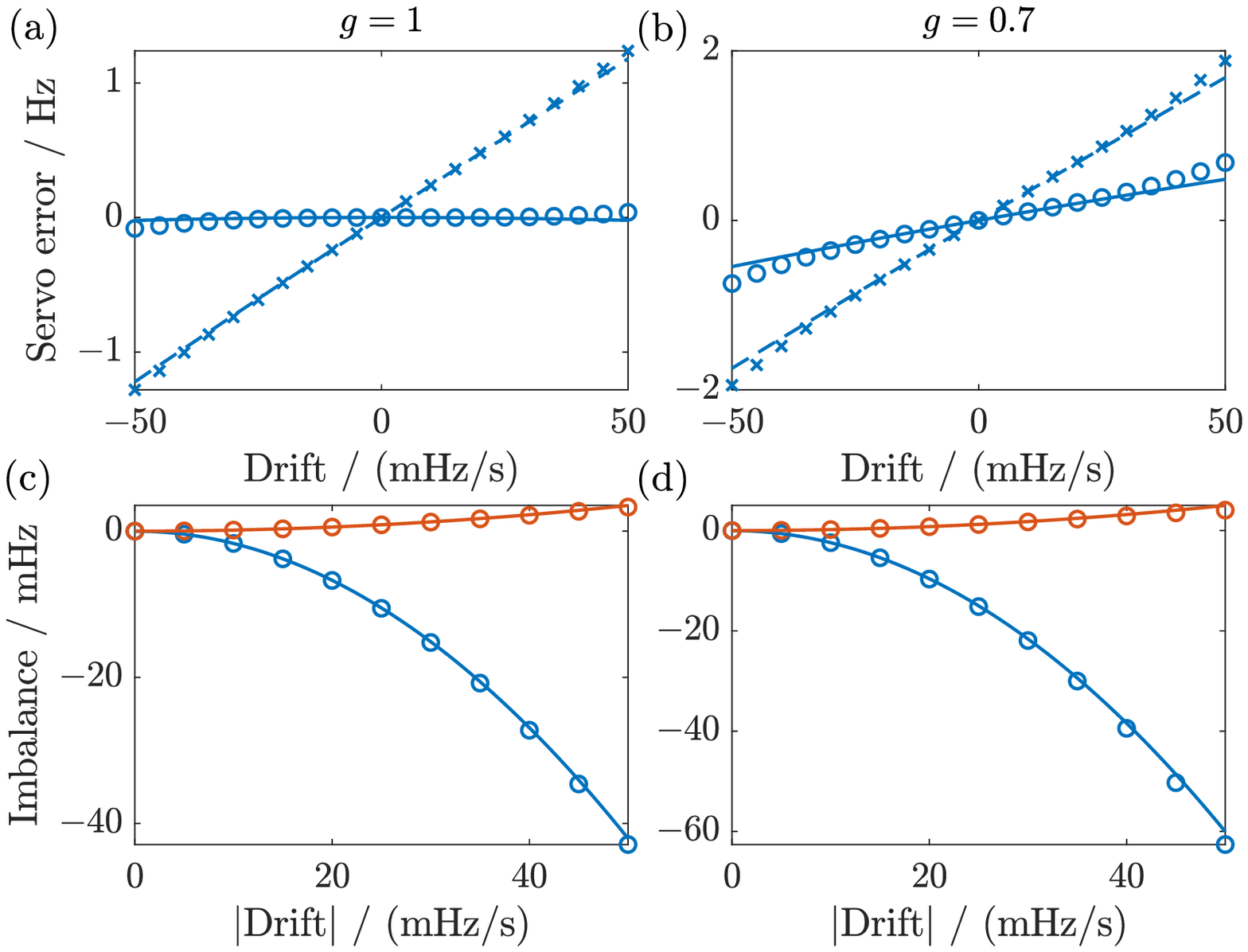}
\caption{(a--b) Real-time (crosses and dashed line) and post-processed (circles and solid line) servo error as a function of the laser drift for the normalized discriminator and gains $g=1$ and $0.7$. (c--d) Imbalance between the error for positive and negative drift as a function of the absolute value of the drift for the same gains.  Blue (red) lines/circles correspond to the normalized (unnormalized) discriminator. Lines are plotted using (\ref{eq:drift-rt}--\ref{eq:drift-pp}), while symbols are from servo simulations.
\label{fig:lin-drift}}
\end{figure}

Figures~\ref{fig:lin-drift}(a--b) show the servo errors (\ref{eq:drift-rt}--\ref{eq:drift-pp}) compared to numerical results from servo simulations (see Section~\ref{sec:sim} for details) for a $^{88}$Sr$^+$ clock, where six transitions are probed and the cycle time is 24\;s. For clarity, only the results for the normalized discriminator are shown; those for the unnormalized one are similar on this scale. For $g=1$, the post-processed servo error is zero in the regime where the discriminator is linear. For the largest drifts, one can see the contribution from the third-order term, which is neglected in the analytical formulas. It could be included in \eqref{eq:pBR}, but solving the servo error for linear drift then requires solving a third-order equation. In practice, a gain $g\approx 1$ is used in combination with another drift-compensation method to reduce servo errors due to fast drift changes \cite{Dube2017a,Dube2018a}. When the real-time frequency is used, this can be achieved by minimizing the cycle time instead.

The imbalance between the servo error for positive and negative drifts is the same for the real-time and post-processed frequencies (in the second-order expansion),
\begin{equation}
\Delta\omega_{\mathrm{rt/pp}}(|d|) + \Delta\omega_{\mathrm{rt/pp}}(-|d|) =
-a' \frac{(|d| T_\mathrm{s})^2}{2 g n_\mathrm{t}}.
\end{equation}
This is  plotted in Figures~\ref{fig:lin-drift}(c--d) and shows good agreement with the simulations.
As the imbalance is caused by the same term in the discriminator that causes the noise-induced servo error, it is relevant to note that the second-order expansion is sufficient to describe the imbalance and that this is much larger for the normalized discriminator.

\section{Ramsey interrogation \label{app:Ramsey}}

First we consider the simplest form of Ramsey interrogation, where two $\pi/2$ pulses are separated by a dark period of duration $T$.
The action of a light pulse with Rabi frequency $\Omega_0$, frequency $\omega_\mathrm{p}$, phase $\varphi$, and duration $\tau$ on a two-level atom with ground state $|g\rangle = (0, 1)\tran$ and excited state $|e\rangle = (1, 0)\tran$ can be expressed using the matrix \cite{Yudin2010a,Meystre-EQO}
\begin{eqnarray}
W(\Omega_0,\delta,\varphi,\tau) = \nonumber \\
\quad \left( \begin{array}{cc}
    \cos{\frac{\Omega \tau}{2}} + i\frac{\delta}{\Omega} \sin{\frac{\Omega \tau}{2}} &
    i\frac{\Omega_0}{\Omega} e^{-i\varphi} \sin{\frac{\Omega \tau}{2}} \\
    i\frac{\Omega_0}{\Omega}  e^{i\varphi} \sin{\frac{\Omega \tau}{2}} &
    \cos{\frac{\Omega \tau}{2}} - i\frac{\delta}{\Omega} \sin{\frac{\Omega \tau}{2}}
    \end{array} \right).
\end{eqnarray}
Here $\delta = \omega_\mathrm{p} -\omega_0$ is the detuning, where $\omega_0$ is the atomic resonance frequency, and $\Omega = (\Omega_0^2 + \delta^2)^{1/2}$ is the generalized Rabi frequency. The free evolution during the dark period is described by \cite{Yudin2010a}
\begin{equation}
V(\delta, T) = \left( \begin{array}{cc}
    e^{i \delta T/2} & 0 \\
    0 & e^{-i \delta T/2}
    \end{array} \right).
\end{equation}
In the simplest scheme, the detuning is constant during the whole Ramsey sequence and the relative phase between the second and first pulses is varied between $+\pi/2$ and $-\pi/2$ to form the discriminator
\begin{eqnarray}
D = |\langle e| W(\Omega_0,\delta_{2+}',\pi/2,\tau) V(\delta_{\mathrm{f}+}', T) W(\Omega_0,\delta_{1+}',0,\tau) |g\rangle |^2 \nonumber \\
-|\langle e| W(\Omega_0, \delta_{2-}',-\pi/2,\tau) V(\delta_{\mathrm{f}-}', T) W(\Omega_0,\delta_{1-}',0,\tau) |g\rangle |^2. \nonumber \\
 { }  \label{eq:D-Ramsey}
\end{eqnarray}
For brevity, we wrote the detunings including the sampled laser noise, $\delta_{i\pm}' = \delta + \delta_{i\pm}$, where the index 1 (2) refers to the first (second) pulse, $\mathrm{f}$ refers to the free evolution, and $\pm$ refers to the sign of the phase step. Assuming the pulse lengths are close to perfect $\pi/2$ pulses, $\Omega_0 \tau = (1+\epsilon) \pi/2$, we can expand \eqref{eq:D-Ramsey} to second order in the detuning and to first order in the small parameter $\epsilon$, which gives the discriminator
\begin{eqnarray}
D = & -\left( \delta + \bar{\delta}_\mathrm{f} \right) T 
 - \left(\frac{4}{\pi} - \frac{2\pi-4}{\pi}\epsilon\right) \left( \delta + \bar{\delta}_\mathrm{p} \right) \tau \nonumber \\
& - \left( \delta \Delta_\mathrm{p} + \frac{\Delta_{2,\mathrm{p}}}{2} \right) \frac{4-\pi}{2\pi} \tau^2 \epsilon. \label{eq:D-Ramsey-s}
\end{eqnarray}
Here, we have defined the mean noise during the free evolution $\bar{\delta}_\mathrm{f} = (\delta_{\mathrm{f}-} + \delta_{\mathrm{f}+})/2$, the mean noise during the pulses $\bar{\delta}_\mathrm{p} = (\delta_{1-}+\delta_{1+}+\delta_{2-}+\delta_{2+})/4$, the differential noise $\Delta_\mathrm{p} = \delta_{1-}-\delta_{1+}+\delta_{2-}-\delta_{2+}$, and the differential squared noise $\Delta_{2,\mathrm{p}} = \delta_{1-}^2-\delta_{1+}^2+\delta_{2-}^2-\delta_{2+}^2$.

For perfect $\pi/2$ pulses, the second-order term in \eqref{eq:D-Ramsey-s} vanishes. For non-perfect pulses, the ratio of the second-order and first-order terms is ${\sim} (4-\pi) \epsilon\tau^2 /(2\pi T -8 \tau)$. The ratio $T/\tau$ is typically on the order of $10\ldots100$. Using the lower value and a 10\% pulse-area accuracy ($\epsilon=0.1$), the ratio of the second-order and first-order terms becomes $1.6\times 10^{-4} T$. This is significantly smaller than the corresponding ratios $a'$ for the normalized and unnormalized Rabi discriminators, which are $0.30 \tau$ and $-0.023 \tau$, respectively, for the ideal lineshape in Figure~\ref{fig:lineshape}.

In hyper-Ramsey or generalized Ramsey spectroscopy~\cite{Yudin2010a,Huntemann2012b}, an additional $\pi$ pulse is added in order to suppress the light shift caused by the clock laser. This makes the expression for the discriminator more complex, but the ratio of the second-order and first-order terms is still on the order of $\epsilon\tau^2 /T$. Even for $T/\tau = 4$ used in the first demonstration~\cite{Huntemann2012b} and $\epsilon=0.1$, this is only ${\sim}6\times 10^{-3} T$. Thus we expect the noise-induced servo error to be negligible in the commonly used Ramsey schemes.

The Ramsey discriminator \eqref{eq:D-Ramsey-s} is steeper than the Rabi discriminators (\ref{eq:Dns}--\ref{eq:Dus}) for the same interrogation time, which results in a lower instability limit \cite{Peik2006a}. The Ramsey scheme is, however, more complex, in particular in clocks with linear Zeeman shift, where interleaved Rabi excitation might be required to ensure locking to the central Ramsey fringe~\cite{Zhang2020c}. The shorter pulses also result in a higher clock-transition light shift, unless this is compensated for using some special technique \cite{Yudin2010a,Huntemann2012b,Sanner2018a}, and can increase the frequency shift and uncertainty due to AOM chirp (optical path length changes caused by radiofrequency heating of the acousto-optic modulator crystal)~\cite{Kazda2016a}.

\section*{References}


\providecommand{\newblock}{}

\end{document}